\RequirePackage{ifpdf}

\documentclass{PoS}
\usepackage[authoryear,square]{natbib}
\bibpunct{(}{)}{;}{a}{}{,}

 \usepackage{graphicx}
 \usepackage{epstopdf}

\title{Cosmology from the EoR/Cosmic Dawn with the SKA}

\ShortTitle{EoR/CD Cosmology}

\author{\speaker{Jonathan Pritchard}$^1$, Kiyotomo Ichiki$^2$, Andrei Mesinger$^3$, Robert Benton Metcalf$^4$, Alkistis Pourtsidou$^4$, Mario Santos$^5$, Filipe Abdalla$^6$, Tzu-Ching Chang$^7$, Xuelei Chen$^8$, Jochen Weller$^{9}$, Saleem Zaroubi$^{10}$,
on behalf of the Cosmology-SWG and EoR/CD-SWG

\\
        $^1$Imperial College London,
        $^2$Nagoya University,
        $^3$Scuola Normale Superiore, Pisa,
        $^4$Universit\'{a} di Bologna,
        $^5$University of the Western Cape,
        $^6$University College London,
        $^7$ASIAA,
        $^8$National Astronomical Observatory of China,
        $^{9}$ Universitaets-Sternwarte Muenchen,
        $^{10}$University of Groningen.
        \\
        E-mail: \email{j.pritchard@imperial.ac.uk}}

\abstract{SKA Phase 1 will build upon early detections of the EoR by precursor instruments, such as MWA, PAPER, and LOFAR, and planned instruments, such as HERA, to make the first high signal-to-noise measurements of fluctuations in the 21 cm brightness temperature from both reionization and the cosmic dawn. This will allow both imaging and statistical maps of the 21cm signal at redshifts $z=6-27$ and constrain the underlying cosmology and evolution of the density field. This era includes nearly 60\% of the (in principle) observable volume of the Universe and many more linear modes than the CMB, presenting an opportunity for SKA to usher in a new level of precision cosmology. This optimistic picture is complicated by the need to understand and remove the effect of astrophysics, so that systematics rather than statistics will limit constraints.

This chapter describes the cosmological, as opposed to astrophysical, information available to SKA Phase 1. Key areas for discussion include: cosmological parameters constraints using 21cm fluctuations as a tracer of the density field; lensing of the 21cm signal, constraints on heating via exotic physics such as decaying or annihilating dark matter; impact of fundamental physics such as non-Gaussianity or warm dark matter on the source population; and constraints on the bulk flows arising from the decoupling of baryons and photons at $z=1000$. The chapter explores the path to separating cosmology from `gastrophysics', for example via velocity space distortions and separation in redshift. We discuss new opportunities for extracting cosmology made possible by the sensitivity of SKA1 and explore the advances achievable with SKA2.
}

\FullConference{
Advancing Astrophysics with the Square Kilometre Array\\
June 8-13, 2014\\
Giardini Naxos, Italy}

\newcommand{\ud}{{\rm d}}
\newcommand{\om}{\Omega_m}
\newcommand{\ob}{\Omega_b}
\newcommand{\odm}{\Omega_{dm}}
\newcommand{\ho}{H_0}
\newcommand{\fnl}{f_{nl}}

\begin{document}

\section{Introduction}

The years since the COBE observations of the CMB have ushered in an age of precision cosmology. Key cosmological parameters have been determined by measurements of the distribution of matter in the Universe through WMAP and Planck observations of CMB anisotropies and large volume galaxy surveys such as SDSS. These surveys have made precision measurements of parameters describing the matter content of the Universe - the baryons $\Omega_b$, dark matter $\Omega_c$, dark energy $\Omega_\Lambda$, radiation $\Omega_r$, and neutrinos $\Omega_\nu$ - and the physics of inflation - via the tilt $n_s$, amplitude $A_s$, running $\ud n_s/\ud\log k$ or the primordial potential power spectrum and $r$ the ratio of tensor-to-scalar modes produced by inflation. These measurements have firmly established a working model of our Universe, known widely as the $\Lambda$CDM model of cosmology. The success of this model is despite our ignorance of the physics of key components, such as dark matter and dark energy, and deviations from the standard model could help refine our understanding.

Despite this precision, measuring model parameters is only the first step towards a deep understanding of the underlying physics. Our ignorance of the nature of the dark matter and the dark energy or how neutrinos acquire mass and what value that mass takes are just two questions that modern cosmology hopes to address. Over the next decade two paths will help shed light on this. The simplest is simply to measure these cosmological parameters ever more precisely and over a wider range of times and scales in the hope of gaining further insights. The exemplar of this is with dark energy, where attempts to measure the redshift evolution of the dark energy density, parameterised by an equation of state $w(z)$, might distinguish a true cosmological constant from more general dark energy or modified gravity. For others there are critical thresholds of precision required to distinguish physical scenarios - for example, measuring the sum of the neutrino masses $M_\nu\lesssim0.1$ would determine the neutrino mass hierarchy. Clearly more precision is a good thing, but it is not the only path forward.

\begin{figure}[htbp]
\begin{center}
\includegraphics[scale=0.45]{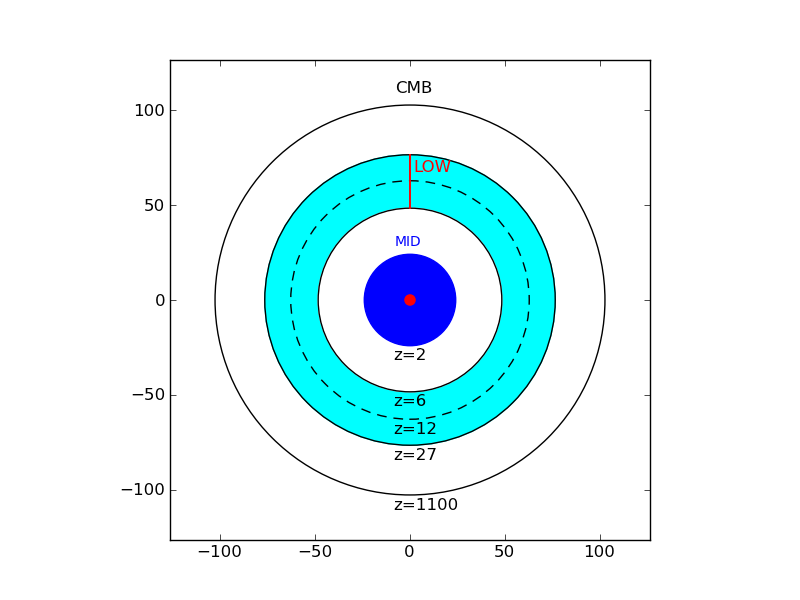}
\caption{Illustration of the volume probed by SKA where 3D comoving volume has been mapped to a 2D disk. The volume probed by an all sky galaxy survey out to $z=0.3$ (red circle, $\sim$SDSS) and $z=2$ (blue circle, $\sim$SKA-MID) are marked as is the volume at the redshifts probed by SKA-LOW, $z=6-27$ (cyan annulus). Existing cosmological parameters are derived using the CMB and relatively local galaxies, with an implicit assumption that nothing strange happens in between.}
\label{fig:volume}
\end{center}
\end{figure}

More generally, we can seek signatures of new physics in ways distinct from the distribution of large scale matter. For example, the processes that produce dark matter will also allow it to annihilate and maybe to decay. The associated release of energy could have impact on the surrounding environment, heating the intergalactic medium. Pursuing unique signatures of new physics in new regimes will be a key part of the next decade.
The SKA is uniquely placed to probe cosmology in this way, as it is capable of mapping the Universe over wide volumes and an unprecedented range of redshifts (see Figure \ref{fig:volume}). In this chapter, we will focus on the new opportunities created by SKA observations of the epoch of reionization (EoR) and the cosmic dawn (CD). This period has never before been observed offering a unique opportunity to test the consistency of the $\Lambda$CDM model and search for new hints to the great unanswered questions of cosmology.

Fundamental physics in cosmology is generally associated with the density field, whose fluctuations are generated by inflation and which contains imprint of other physics such as neutrino mass. Astrophysics is a major challenge to getting at cosmology with SKA, but we can identify several key approaches to extracting cosmology: (1) directly from density fluctuations (2) via the presence of exotic sources of radiation (3) via the radiation fields produced by all sources since those sources will trace the density field in some biased fashion (4) via the weak lensing of the 21cm signal by structures between the observer and signal (5) miscellaneous other probes. Extracting cosmology from the 21cm signal during the EoR will require innovative new techniques to separate astrophysics from cosmology. Nonetheless the sensitivity of the instruments, large volume probed, and new redshift regime accessible to SKA makes this a very interesting area for new science. 

In this chapter, we will explore these different avenues for extracting cosmology from the 21cm signal and attempt to assess the sort of constraints that will be achievable by SKA Phase 1 and 2. However, we caution the reader that this is not a settled area and it is still unclear how well astrophysics can be dealt with. New ideas may improve the constraints, but new obstacles may render them optimistic.

\section{Cosmological parameters from density fluctuations}

In this section, we explore the ability of SKA to constrain cosmological parameters via observations of the density field. Just as galaxy surveys constrain cosmology by using galaxies as a tracer of the linear density field, SKA can constrain cosmology by using the 21 cm brightness temperature as a tracer of the density field. This is not an unproblematic assertion, since brightness temperature fluctuations may be sourced by variation in the spin temperature and neutral fraction in addition to the density field. 

\begin{equation}\label{deltatb}
\delta T_B=27x_{HI}(1+\delta_b)\left(\frac{T_S-T_{\rm CMB}}{T_S}\right)\left(\frac{1+z}{10}\right)^{1/2}\left[\frac{\partial_rv_r}{(1+z)H(z)}\right]^{-1}{\rm\,mK}
\end{equation}
Equation \ref{deltatb} shows how these different terms come into play \citep{2006PhR...433..181F}. In a regime where $T_S\gg T_{\rm CMB}$ and $x_{HI}=1$ then $\delta T_B$ will be an unbiased tracer of the density field. At all other times the effects of astrophysics must be modelled and removed or somehow avoided. One possibility might be to exploit redshift space distortions that produce an angular dependence of the power spectrum, which in the simplest linear theory models look like
\begin{equation}
P(k)=P_{\mu^0}(k)+P_{\mu^2}\mu^2+P_{\mu^4}\mu^4.
\end{equation}
In principle, measurement of this angular dependence of the power spectrum could separate cosmology and astrophysics since the $P_{\mu^4}=P_\delta$ so directly probes the density field. In practice, this separation is complicated by non-linear growth of structure \citep{2008PhRvD..78j3512S, 2012MNRAS.422..926M} and the motion of ionised regions themselves \citep{2006ApJ...653..815M} and it is unclear how effective it can be. We will return to a discussion of separating astrophysics and cosmology in \S\ref{sec:separation} as this is a critical point. 

In this section, we take the optimistic view that there will a regime in which $\delta T_b\propto(1+\delta)$ so that the 21cm signal provides a clean measurement of the density field. This approach enables us to evaluate the best case scenario for SKA in measuring cosmological parameters. By comparing this to galaxy surveys we get a sense of how competitive SKA could be, if astrophysics could be overcome.
While we focus on standard cosmological parameters - energy density in baryons $\Omega_b$, matter $\Omega_m$, cosmological constant $\Omega_\Lambda$; hubble parameter $h$; inflationary parameters $A_S$, $n_s$ and $\ud n_s/\ud\log k$; neutrino mass $M_\nu$ and curvature $\Omega_k$ - SKA will open a new regime into exotic physics that can only be probed at high redshift, for example compensated isocurvature modes whose effect decreases with time \citep{2009PhRvD..80f3535G}. Cosmology is moving from simply wanting to measure cosmological parameters more accurately and instead becoming more focused on control of systematics and relaxing simplifying assumptions. SKA will test consistency of cosmological parameters in a new redshift range.

The sensitivity of a radio interferometer to the 21cm power spectrum has been well studied \cite[e.g.][]{bowman2006, 2006ApJ...653..815M,2008PhRvD..78b3529M,2013ExA....36..235M} and we follow the same approach here. The variance of a 21 cm power spectrum estimate for a single
$\mathbf{k}$-mode with line of sight component $k_{||}=\mu k$ is given by \citep{2008ApJ...680..962L}:
\begin{equation}
\sigma_P^2(k,\mu)= \frac{1}{N_{\rm field}}\left[\bar{T}_b^2P_{21}(k,\mu)+T_{\rm sys}^2\frac{1}{B t_{\rm int}}\frac{D^2\Delta D}{n(k_\perp)}\left(\frac{\lambda^2}{A_e}\right)^2\right]^2.
\end{equation}

The first term on the right-hand-side
of the above expression provides the contribution from sample variance,
while the second describes the thermal noise of the radio telescope.  The
thermal noise depends upon the system temperature $T_{\rm sys}$, the survey
bandwidth $B$, the total observing time $t_{\rm int}$, the conformal
distance $D(z)$ to the center of the survey at redshift $z$, the depth of
the survey $\Delta D$, the observed wavelength $\lambda$, and the effective
collecting area of each antennae tile $A_e$.  The effect of the
configuration of the antennae is encoded in the number density of baselines
$n_\perp(k)$ that observe a mode with transverse wavenumber $k_\perp$
\citep{2006ApJ...653..815M}.  Observing a number of fields $N_{\rm field}$ further reduces the variance. Given the sensitivity of the instrument, the Fisher matrix formalism can be used to estimate $1-\sigma$ errors on the model parameter $\lambda_i$ are $(\mathbf{F}_{ij}^{-1})^{1/2}$, where 
\begin{equation}
F_{ij}=\sum_{\rm \mu} \frac{\epsilon k^3 V_{\rm survey}}{4\pi^2}\frac{1}{\sigma_P^2(k,\mu)}\frac{\partial P_{T_b}}{\partial \lambda_i}\frac{\partial P_{T_b}}{\partial \lambda_j}.
\end{equation}
In this equation, $V_{\rm survey}=D^2\Delta D(\lambda^2/A_e)$ denotes the
effective survey volume of our radio telescopes and we assume wavenumber
bins of width $\Delta k=\epsilon k$.  Key to determining cosmological parameters are the effective volume probed and the minimum wavenumber probed $k_{\rm min}$ where modes can still be assumed to be linear. SKA has a significant advantage over galaxy surveys as more modes are still in the linear regime at $z>6$. 

\begin{table}[htdp]
\caption{Low-frequency radio telescopes and their parameters.  We specify the number of antennae $N_a$, total collecting area $A_{\rm tot}$, bandwidth $B$, and total integration time $t_{\rm int}$ for each instrument. These values are fixed at $\nu=110$MHz and extrapolated to other frequencies using $A_{\rm tot}=N_{\rm ant}N_{\rm dip}A_{\rm dip}$ with a physical station size of 35m and the number of antennae per station $N_{\rm dip}=289$ and $A_{\rm dip}=\min(\lambda^2/3,3.2\,{ \rm m^2})$.}
\begin{center}
\begin{tabular}{ccccccc}
\hline
\hline
Array & $N_a$ & $A_{\rm tot}(10^3\,{\rm m^2})$ & $B$ (MHz) & $t_{\rm int}$ (hr)& $R_{\rm min} (m)$ & $R_{\rm max} (km)$\\
\hline
MWA & 112 & 1.6  & 8 & 1000 & 4 & 0.75\\
PAPER & 128 & 0.9  & 8 & 1000 & 4 & 0.15\\
LOFAR Core & 48 & 38.6  & 8 & 1000 & 100 & 1.5\\
HERA & 331 & 50.0  & 8 & 1000 & 14.3 & 0.3\\
SKA0 & 899$\times$0.5 & 831$\times$0.5  & 8 & 1000 & 35 & 2\\
SKA1 & 899 & 831  & 8 & 1000 & 35 & 2\\
SKA2 & 899$\times$4 & 831$\times$4 & 8 & 1000 & 35 & 2\\
\hline
\hline
\end{tabular}
\end{center}
\label{tab:telescopes}
\end{table}%

\begin{figure}[htbp]
\begin{center}
\includegraphics[scale=0.35]{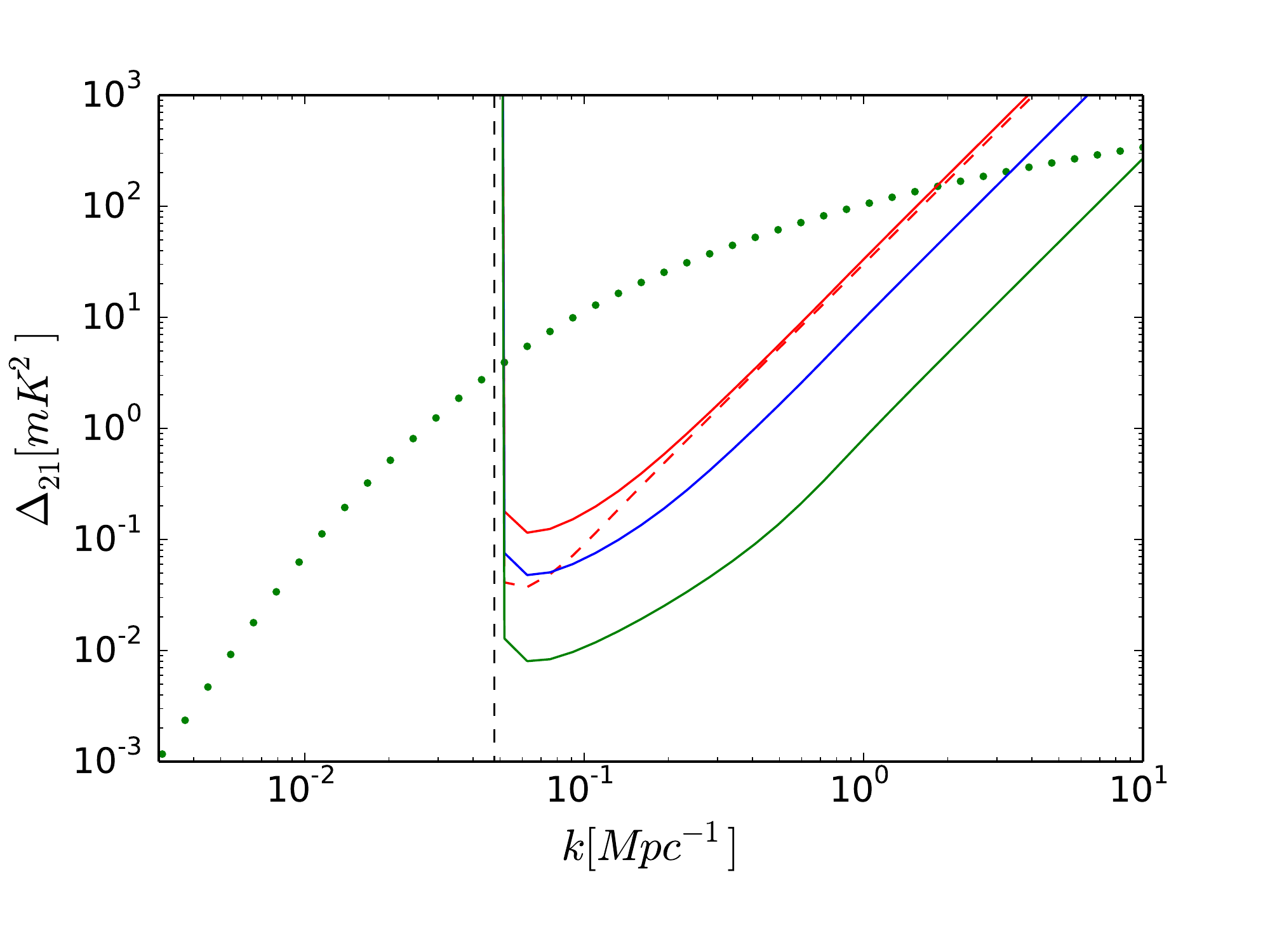}
\includegraphics[scale=0.35]{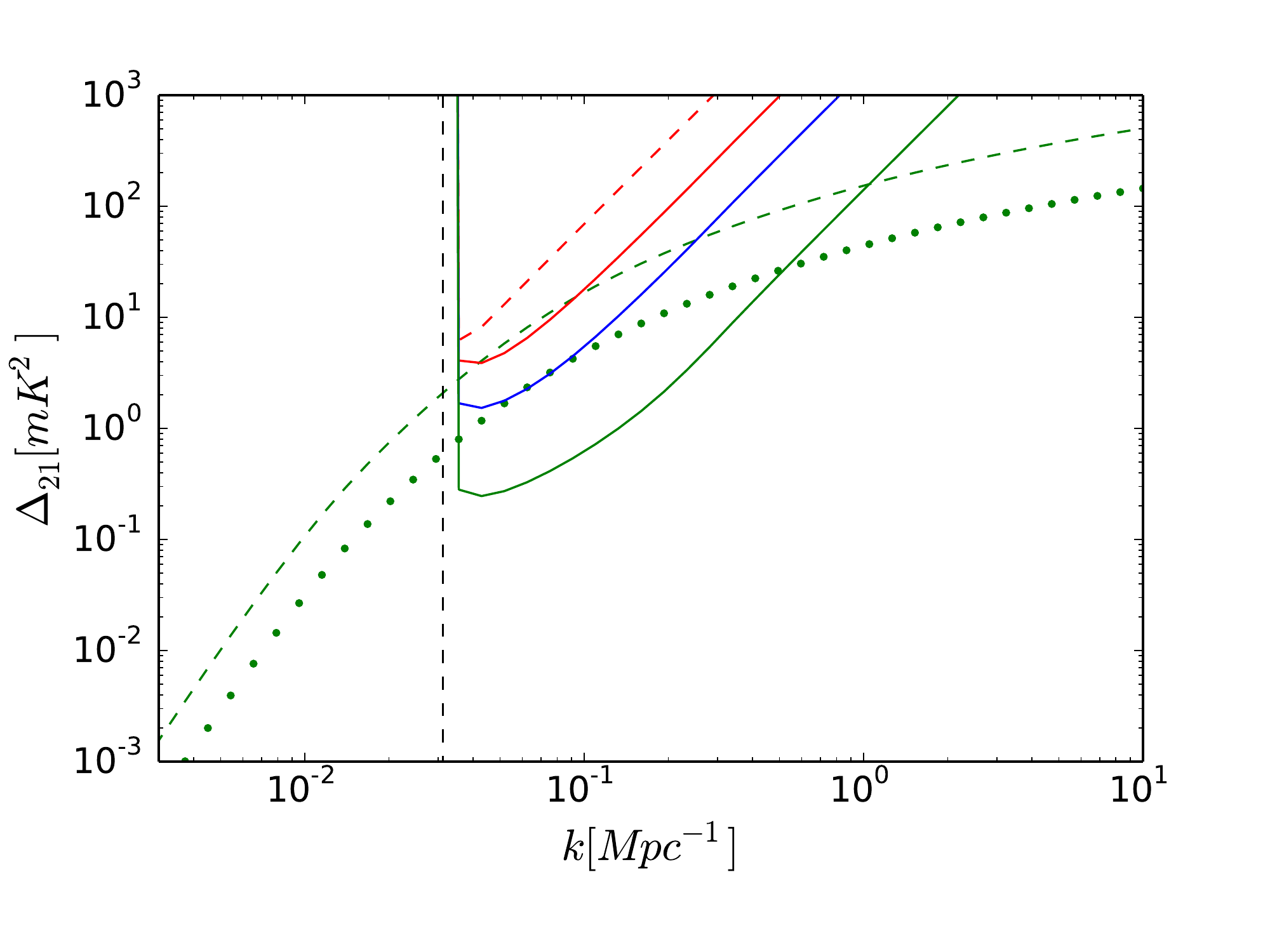}
\caption{Sensitivity plots of HERA (red dashed curve), SKA0 (red), SKA1 (blue), and SKA2 (green). Dotted curve shows the predicted 21cm signal {\em from the density field alone} assuming $x_H=1$ and $T_S\gg T_{\rm CMB}$. At $z=20$, we also plot the case of $T_S=20{\rm K}$ in the $z=20$ panel to give a better sense of the expected 21 cm signal during absorption. Vertical black dashed line indicates the smallest wavenumber probed in the frequency direction $k=2\pi/y$, which may limit foreground removal.  {\em Left panel:} $z=8$ {\em Right panel:} $z=20$.}
\label{fig:sensitivity}
\end{center}
\end{figure}

We first illustrate the sensitivity of different iterations of SKA in Figure \ref{fig:sensitivity}, where we take the parameters in Table \ref{tab:telescopes} for SKA0 - with 50\% of the SKA1 baseline collecting area, SKA1, and SKA2 - with x4 the collecting area of SKA1. For each of these we assume a filled core followed by $r^{-2}$ distribution out to a maximum radius $R_{\max}$. HERA is assumed to have a uniform antennae distribution. SKA1 has 911 stations total with 899 in the core and 650 stations within a radius of 1km accounting for $\sim$75\% of the total number of stations and collecting area \citep{Dewdney:2013}. At lower frequencies the array is densely packed and has constant collecting area, at higher frequencies the array becomes sparse.

Figure \ref{fig:sensitivity} illustrates a few key points governing parameter constraints. Here we have eliminated modes whose wavelength exceeds the instrument bandwidth removing sensitivity to the largest physical scales (smallest $k$ modes). At $z=8$, SKA0 is directly comparable in sensitivity to the proposed HERA experiment \citep{2014ApJ...782...66P}, which is more centrally concentrated to compensate for its small number of stations. Detection of the 21cm signal at $z\gtrsim20$ with SKA1 is dependent upon either a strong 21cm absorption signal that boosts the amplitude of the 21cm power spectrum, e.g. $T_S\ll T_\gamma$ as expected before X-ray heating, or spin temperature fluctuations that add additional power over that of the density field. Unfortunately, it seems likely that during the absorption regime the details and spatial variation of the spin temperature will matter and complicated getting at cosmology.

Table \ref{tab:constraints} shows the cosmological parameters obtained with the listed experimental performances using a Fisher matrix approach following \citet{2006ApJ...653..815M}. The key take home message of this is that SKA-LOW has the raw sensitivity to add useful information on cosmological parameters to Planck. The largest gains are on parameters that require small scale information, for example the running of the primordial power spectrum and related inflationary parameters \citep{2009PhLB..673..173B,2011JCAP...02..021A} and neutrino masses \citep{2008PhRvD..78f5009P}. This also indicates that SKA-LOW will have the sensitivity to provide a useful consistency check on cosmological parameters from the high redshift regime long before dark energy becomes important.

These numbers assume a single deep field designed to reduce thermal noise and so maximise sensitivity on the smallest scales. This tends to maximise the constraint on parameters like neutrino mass, which modify the power spectrum primarily on small scales. On large scales, cosmic variance dominates over thermal noise. This makes it useful to complement a single deep field with many shallower fields, which increase the survey volume and reduce the cosmic variance. The SKA-LOW survey strategy of shallow $\sim10000$ deg$^2$, mid $1000$ deg$^2$, and deep $100$ deg$^2$ surveys provides a good mix to optimise for cosmology.

\begin{table*}[htdp]
\caption{Fiducial parameter values and $1-\sigma$ constraints on cosmological parameters. Non-cosmological parameters included in the analysis \{$\tau$, $x_H(z=7)$, $x_H(z=7.5)$, $x_H(z=8)$\} are not shown. We take $k_{\rm min}=2{\rm Mpc^{-1}}$ as the limit to linear modes.}
\begin{center}
\begin{tabular}{c|cccccccc}
\hline
 & $\log\Omega_mh^2$ & $\log\Omega_bh^2$ & $\Omega_\Lambda$ & $n_s$ & $\log (A_s/10^{-10})$ & $\Omega_k$ & $dn_s/d\log k$ & $M_\nu$ (eV) \\
Value & -1.9 & -3.8 & 0.7 & 0.95 & -0.19 & 0 & 0 & 0.3\\
\hline
Planck & 0.028 & 0.0068 & 0.038 & 0.0035 & 0.0097 & 0.0022 & 0.0047 & 0.35 \\
Hera & 0.0091 & 0.0055 & 0.011 & 0.003 & 0.0088 & 0.0021 & 0.0036 & 0.12  \\
SKA0 & 0.017 & 0.0058 & 0.023 & 0.0032 & 0.009 & 0.0022 & 0.0034 & 0.22  \\
SKA1 & 0.0083 & 0.0051 & 0.01 & 0.003 & 0.0084 & 0.002 & 0.0018 & 0.12  \\
SKA2 & 0.0016 & 0.0048 & 0.0026 & 0.0027 & 0.0081 & 0.0012 & 0.00092 & 0.084 
\end{tabular}
\end{center}
\label{tab:constraints}
\end{table*}

We make no attempt here to model the effects of astrophysics on these constraints. Increasingly conservative assumptions can degrade these constraints arbitrarily far \citep{2008PhRvD..78b3529M}, so these should be viewed as optimistic bounds on the constraints that might be achieved. Nonetheless it is clear that the attempt to extract cosmology from CD and EoR observations could be quite rewarding.

\section{Separating ``gastrophysics" and cosmology}
\label{sec:separation}

\begin{figure}[htbp]
\begin{center}
\includegraphics[scale=0.45]{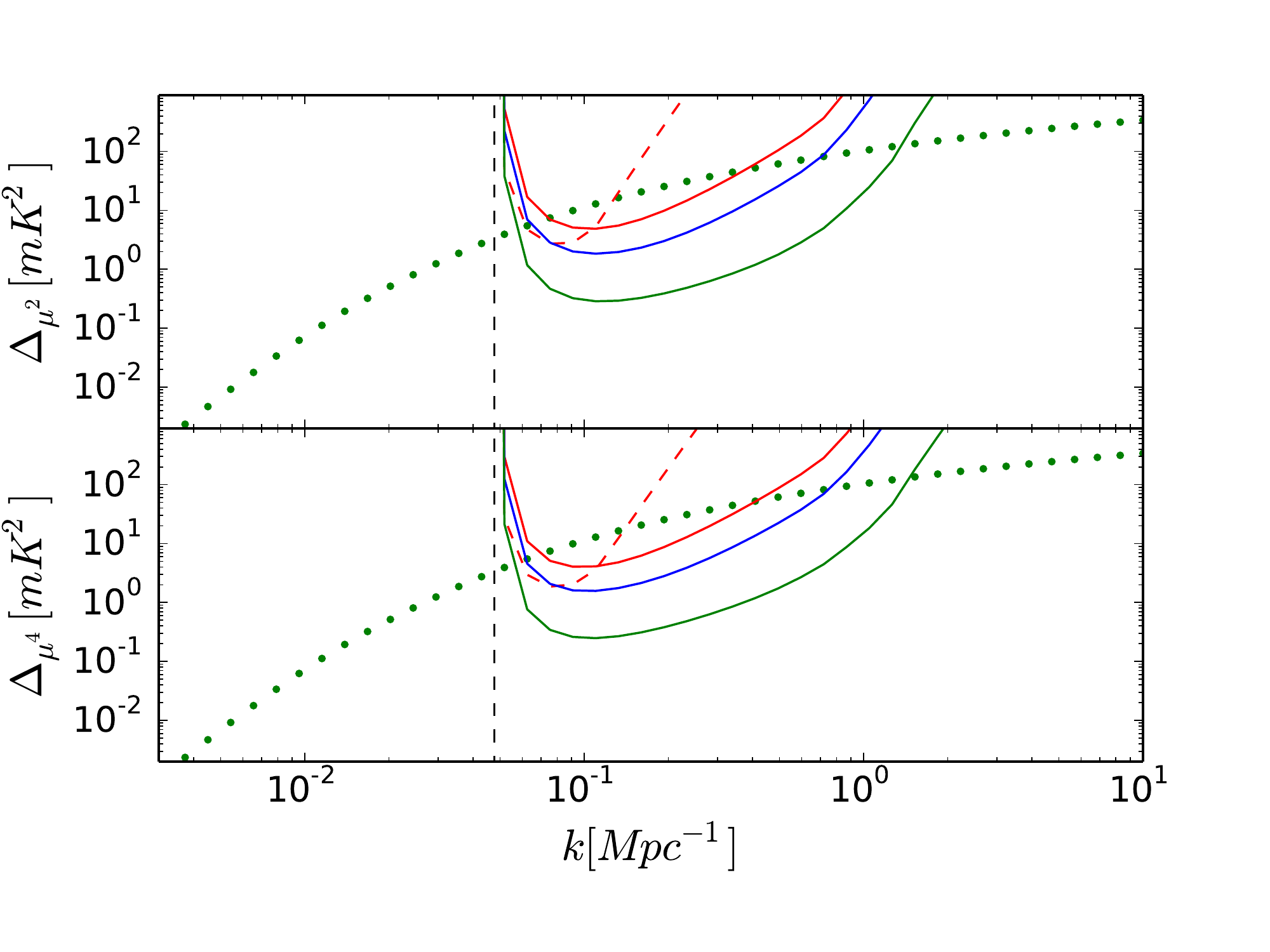}
\caption{Sensitivity plots at $z=8$ on $P_{\mu^2}$ (top panel) and $P_{\mu^4}$ (bottom panel) for HERA (red dashed curve), SKA0 (red), SKA1 (blue), and SKA2 (green). Dotted curve shows the predicted 21cm signal from the density field alone assuming $x_H=1$ and $T_S\gg T_{\rm CMB}$. Vertical black dashed line indicates the smallest wavenumber probed in the frequency direction $k=2\pi/y$, which may limit foreground removal. }
\label{fig:musensitivity}
\end{center}
\end{figure}

The key challenge for extracting fundamental physics from the 21cm signal will be separating the effects of cosmology from ``gastrophysics". A number of avenues have been studied in the literature, which broadly separate into (1) avoidance and (2) modelling and (3) redshift space distortions. The optimistic case in the previous section assumed the possibility of avoidance - a clean region in redshift where $T_S\gg T_{\rm CMB}$ and $x_H=1$. Theoretical modelling of the 21cm signal \citep[e.g.][]{2008PhRvD..78j3511P,2011MNRAS.410.1377T,2013MNRAS.431..621M,2014MNRAS.437L..36F} suggests that we are unlikely to find such a region, although certain epochs may approach this limit. In the absence of a clean window, it might still be possible to avoid astrophysics via the angular dependence of the power spectrum induced by redshift space distortions. Focussing on the $P_{\mu^4}\approx P_\delta$ part could lead to clean cosmological measurements. Obtaining precision cosmology this way is hard and the literature suggests little improvement over Planck will be possible \citep{2006ApJ...653..815M,2008PhRvD..78b3529M}. Figure \ref{fig:musensitivity} shows predicted errors bars for SKA on the $P_{\mu^2}$ and $P_{\mu^4}$ parts of the power spectrum. A detection is possible with at wavenumbers $k=0.1-1{\rm\,Mpc^{-1}}$, but with much less precision than the full 21 cm power spectrum. 

The most likely path is to model the contribution of astrophysics. Compared with the CMB our theoretical understanding of the 21cm signal during reionization is poor. Predictions for the 21cm power spectrum do not exist at the same level of precision as the cosmology. Nonetheless, we expect the contribution of astrophysics to be relatively broad band and determined by extra power about a characteristic scale, eg the bubble size during reionization. \cite{2008PhRvD..78b3529M} showed that relatively simple parametrisations capture the shape of ionisation contributions and so might be fitted for and marginalised out. Information from measurements of $P_{\mu^2}$ would complement this, as would information from different redshift slices. Given the large amount of information in 3D and the ability of SKA to image the signal - allowing ionised bubbles to be directly identified and masked out - it may be possible to characterise the astrophysics on large scales. This has yet to be examined in detail and it is unclear how far 21cm observations might be ``cleaned" of astrophysics. 

One thing to note is that reionization destroys information - ionised bubbles produce no 21cm signal - while heating and Ly$\alpha$ coupling merely overlay the density field with other information. It may be possible in the future to establish a way of separating spin temperature fluctuations from the density field in some other way, e.g. by using galaxy observations to reconstruct the Ly$\alpha$ flux on large scales, and so recover more of the cosmological information from observations in the cosmic dawn.

\section{Constraining new physics from heating}
\label{sec:heating}

The 21cm signal probes both the ionization and thermal state of the IGM.  Although we do not know the precise timing and evolution of the signal, empirical scaling relations based on local star-forming galaxies \cite[e.g.][]{2012MNRAS.419.2095M} suggest that the X-rays from early galaxies heat the IGM to temperatures above the CMB before the bulk of reionization \cite[e.g.][]{2006MNRAS.371..867F,2012ApJ...760....3M}.  This marks the transition of the 21cm signal from absorption to emission, with large-scale fluctuations in gas temperature likely driving the 21cm power to its largest amplitude \cite[e.g.][]{2007MNRAS.376.1680P,2010A&A...523A...4B}.  The epoch of IGM heating is a powerful probe of the high-energy processes in the early Universe, with could have both astrophysical and cosmological origins.  Both can tell us about the nature of dark matter (DM).

In order to explain the apparent deficiencies of CDM on small (sub-Mpc) scales, Warm Dark Matter (WDM) models have recently gained in popularity.  In these models, DM is assumed to consist of smaller mass particles, $\sim$ keV, such as the sterile neutrino or gravitino.  The increased particle free-streaming and velocity dispersion (acting as a sort of effective pressure), can dramatically suppress structures on small-scales.  This suppression is even more obvious in the early Universe, where typical halos hosting galaxies were much smaller, and larger structures did not have time to fragment.  Current astrophysical lower limits on the WDM particle range from $m_x \gtrsim$ 1-3 keV (assuming a thermal relic relativistic at decoupling), with various degrees of astrophysical degeneracy \cite[e.g.][]{2013MNRAS.432.3218D,2013ApJ...767...22K,2014MNRAS.443..678P,2013PhRvD..88d3502V}

\begin{figure}[htbp]
\begin{center}
\includegraphics[scale=0.4]{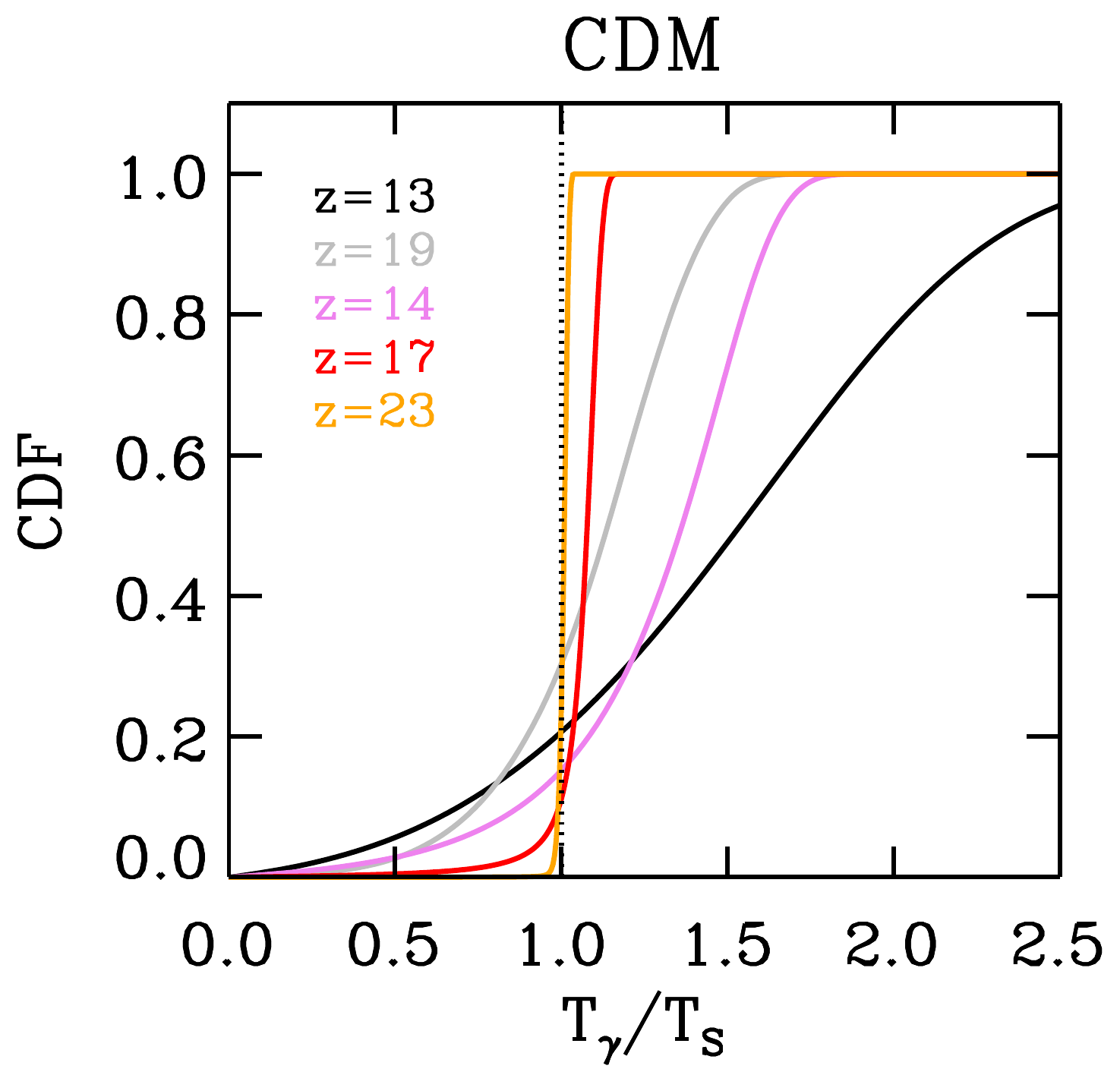}
\includegraphics[scale=0.9]{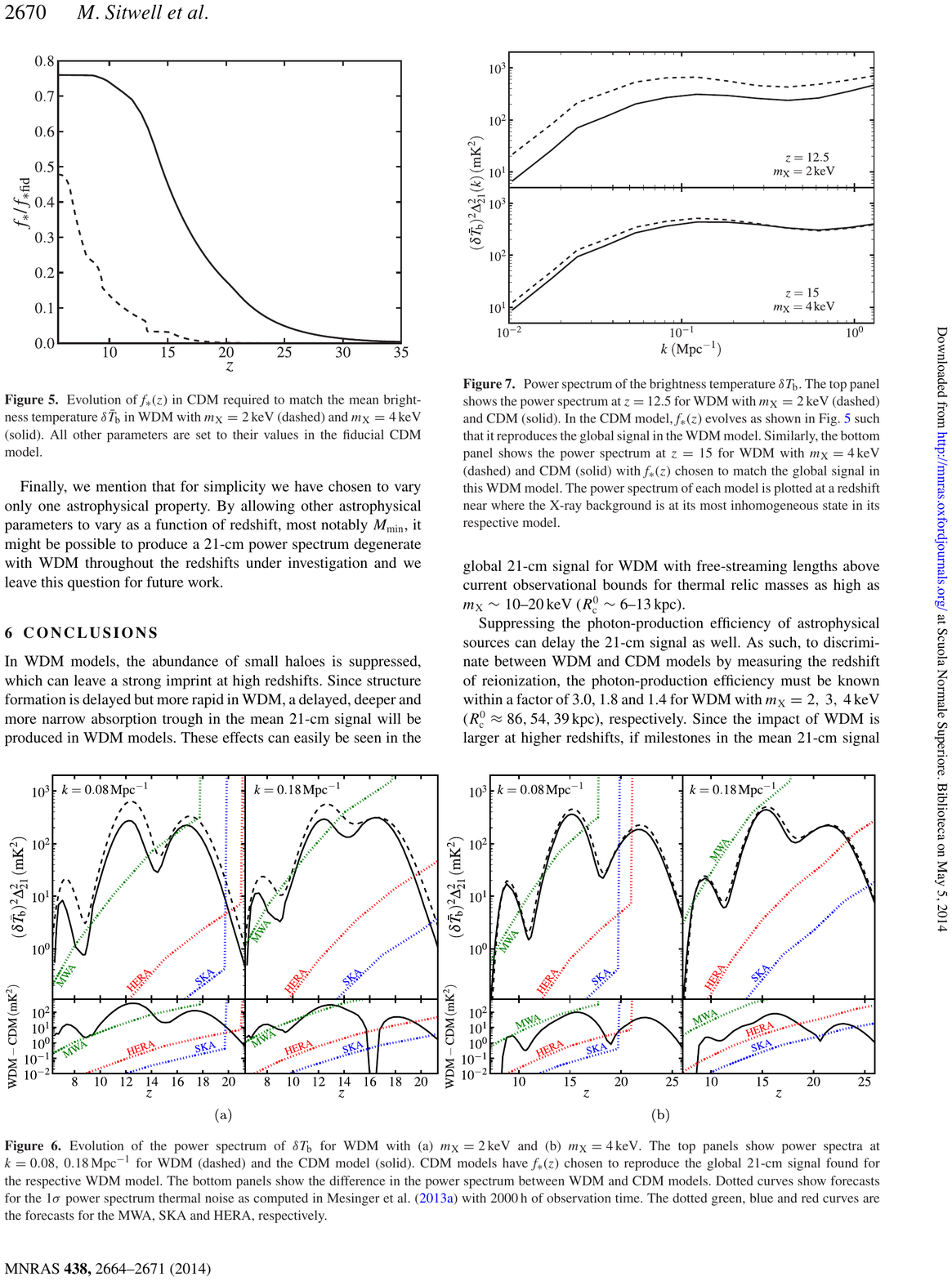}
\caption{{\em Right panel: }Evolution of the power spectrum of $\delta T_b$ for WDM with $m_X = 2$ keV. The top panels show power spectra at
$k =$ 0.08, 0.18 Mpc$^{-1}$ for WDM (dashed) and the CDM model (solid). CDM models have $f_*$(z) (star-formation efficiency) chosen to reproduce the global 21-cm signal found for
the respective WDM model. The bottom panels show the difference in the power spectrum between WDM and CDM models. Dotted curves show forecasts
for the 21cm power spectrum thermal noise as computed in \cite{2014MNRAS.439.3262M} assuming 2000 h of observation time. The dotted green, blue and red curves are
the forecasts for the MWA, SKA and HERA, respectively.  This figure is from \cite{2014MNRAS.438.2664S}.
CDFs of $T_\gamma/T_S$ corresponding to the fiducial and extreme astrophysical X-ray heating (black and gray curves respectively) from \cite{2013MNRAS.431..621M}.  The colored curves correspond to models in which 10 GeV DM annihilations are also accounted for (in addition to fiducial astrophysical heating), with varying relative contribution. The curves correspond to the redshift for which $T_s \sim T_{\rm CMB}$. Figure is from \citep{2014arXiv1408.1109E}.}
\label{fig:darkmatter}
\end{center}
\end{figure}

The resulting dearth of galaxies in the early Universe means that the astrophysical epochs in the 21cm signal were delayed.  The challenge as always will be to disentangle the cosmological impact from astrophysical uncertainties, for example a lower than expected star formation efficiency in CDM would look superficially similar to a higher star formation efficiency in WDM.  Since the fractional suppression of structure increases with redshift, this becomes much easier with the first galaxies observable with the SKA.  For example, we only need to understand the astrophysics of the first galaxies to an order of magnitude in order to improve on current $m_X$ constraints \citep{2014MNRAS.438.2664S}.
Moreover, even if the star-formation efficiency in CDM is allowed to vary in order to mimic the mean 21cm evolution in WDM models, the signal will still not be completely degenerate (see Fig. \ref{fig:darkmatter}a).  This is due to the fact that the galaxies driving the 21cm evolution in WDM should reside in higher mass, more rapidly evolving halos, than those in CDM.  The increased bias of such halos results in a larger 21cm fluctuations (see Fig. \ref{fig:darkmatter}a).

The heating of the IGM could also have a cosmological component.  In particular, annihilations of dark matter particles in the $\sim$ 10 GeV mass range (motivated by recent results from indirect experiments; \cite[e.g.][]{2009Natur.458..607A, 2010JCAP...04..014A,2013PhRvL.110n1102A}
could provide a dominant source of heat, before the birth of the first galaxies.  Driven by the evolution of $\sim M_\odot$ structures, several orders of magnitude smaller than those hosting galaxies, heating is expected to be much slower in such models, resulting in a smaller brightness temperature gradient  $\ud\delta T_b/\ud\nu \sim4 {\rm mK\, MHz^{-1}}$ in the range $\nu \sim 60 - 80$ MHz \citep{2013MNRAS.429.1705V}.  Moreover, DM annihilations would heat the IGM quite uniformly, which is not the case for heating driven by astrophysical sources residing in early galaxies.  The resulting lack of temperature fluctuations (see Fig. \ref{fig:darkmatter}b) would result in dramatic drop in 21cm power during heating, which would be easy to identify with the SKA \citep{2014arXiv1408.1109E}.  Furthermore, the ensuing rise in 21cm power when the galaxies start contributing to heating the IGM should occur {\em when the IGM is already in emission}.  The later is a qualitatively robust signature of DM annihilation heating, easily obtainable with the SKA.

\subsection{21 cm signal from primordial magnetic fields}

Primordial magnetic fields (PMFs) has been intensively investigated in
the literature as possible seeds for large scale magnetic fields
observed in galaxies and clusters of galaxies (for a recent review, see
\citep{2013A&ARv..21...62D}).  Magnetic fields in galaxies in high
redshifts \citep{2008Natur.454..302B} and in void regions
\citep{2010Sci...328...73N,2010ApJ...722L..39A,2013ApJ...771L..42T} can
well be the pieces of evidence that the seed fields are of primordial
origin.  The primordial magnetic fields may be created in the very
early universe, e.g., at the epoch of inflation, cosmological phase
transition, and cosmological recombination.  
The Planck collaboration
recently placed limits on the magnetic field strength smoothed on 1 Mpc scales $B_{\lambda=1{\rm Mpc}} < 3.4$ nG and the spectral index $n_B<0$ of any PMFs
from the temperature anisotropies on large and small angular scales
\citep{2013arXiv1303.5076P}. 

The CMB brightness temperature fluctuations produced by the neutral
hydrogen 21-cm line (21 cm) would offer a new probe of the primordial
magnetic fields (PMFs) created in the early universe. For the 21 cm
observation, aside from the early structure formation effect by the
Lorentz force from the PMFs, one of the important effects is the
dissipation process of the PMFs that increases the baryon
temperature. The dissipation occurs mainly through ambipolar
diffusion due to the velocity difference between neutral hydrogen (which
is the dominant component in the dark ages) and ionized particles (whose
trajectory is bent by the Lorentz force).  The effect of the dissipation
is rather significant. The gas temperature can reach $1000$ K or even
$10^4$ K at $z=30$ if the magnetic fields have the strength of
$B_\lambda \sim 3$ nG
\citep{2005MNRAS.356..778S,2006MNRAS.372.1060T,2009ApJ...692..236S,2014JCAP...01..009K}.

\begin{figure}[]
\includegraphics[width=0.5\linewidth,angle=0]{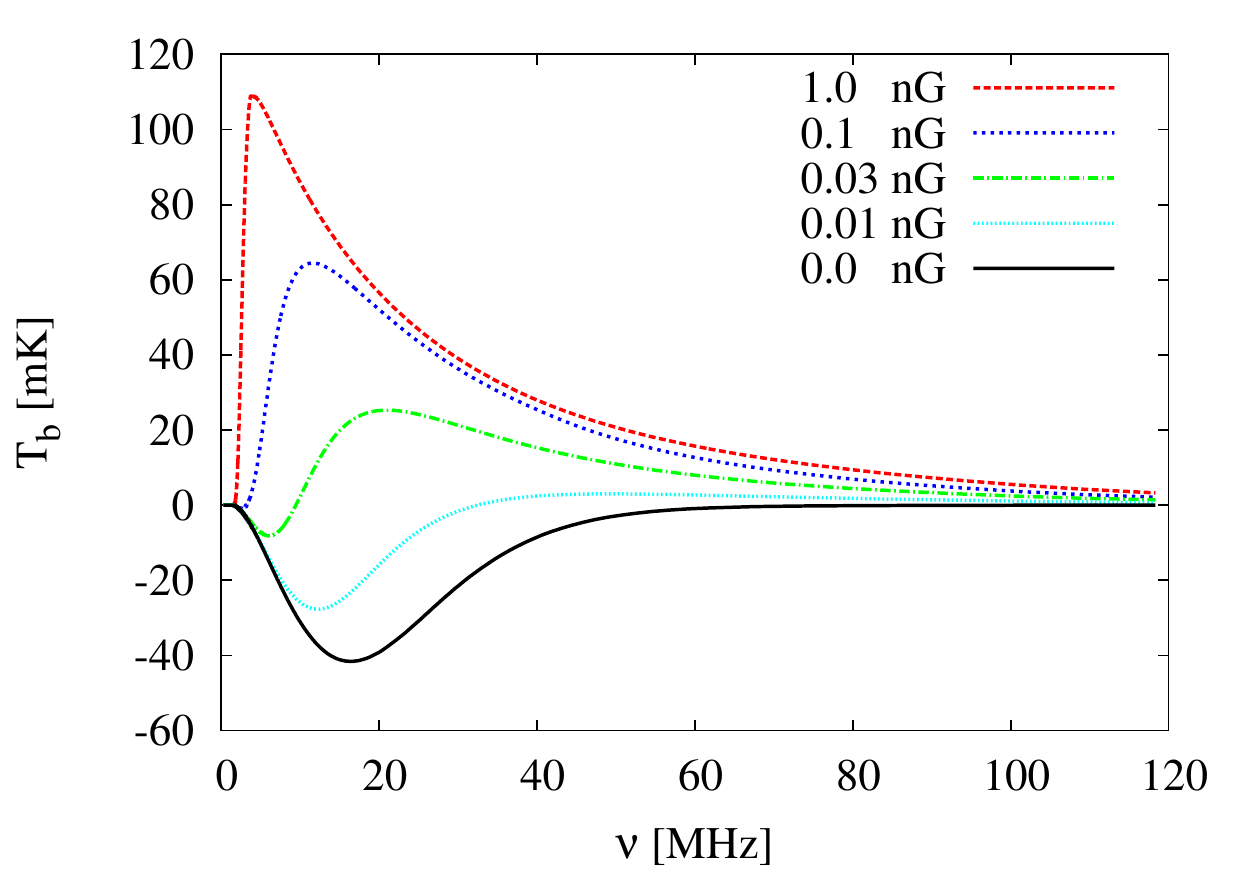}
\includegraphics[width=0.5\linewidth,angle=0]{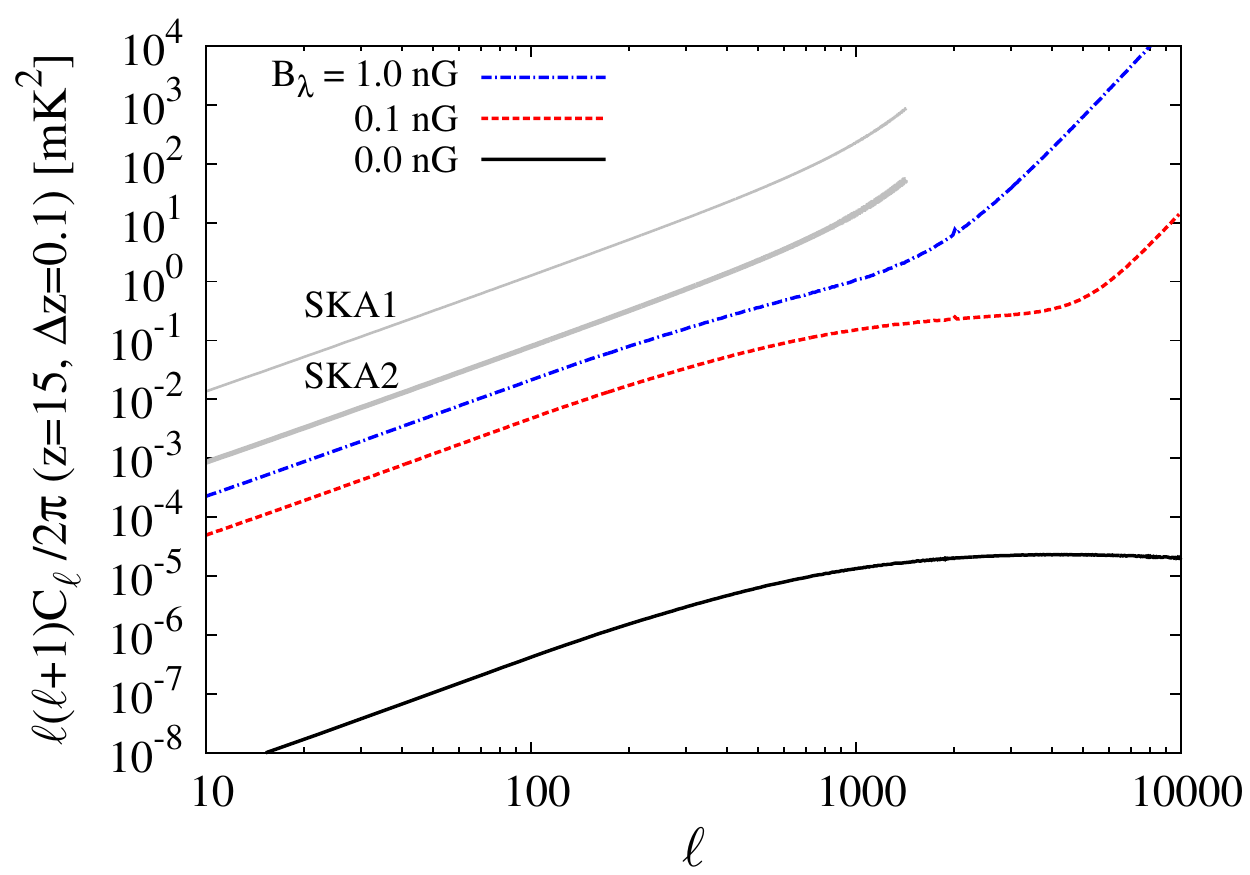}
\caption{{\em Left panel: }The global 21 cm signal with magnetic field strength $B=1$
$0.1$, $0.03$, and $0.01$ nG (colored lines from top to bottom). The
solid line corresponds to the model without primordial magnetic fields.
Note that any other heating source than magnetic fields is neglected in
the figure. {\em Right panel: }Angular power spectra for PMF strengths: $B=0,~ 0.1,~ 1.0$ nG
at $z=15$. The bottom curve shows the power spectrum from the standard
density perturbations for fully neutral medium without any heating and
reionization processes. The red and blue curves correspond to the cases
with heating by the PMFs with $B=0.1$ nG and $B=1.0$ nG,
respectively. The heating induces deviations of the spin temperature
from the CMB temperature and the signal is enhanced. The noise curves
for SKA1 and SKA2 with 1MHz bandwidth are also shown as indicated. By courtesy of
M. Shiraishi \& H. Tashiro. For reference, at $z=15$, $k\approx0.2(l/1000){\rm\,Mpc}^{-1}$.}  \label{fig:KI1}
\end{figure}

This dissipation will give rise to a unique signature of the PMFs on the
21 cm observation. Because the spin temperature is closely coupled to
the gas temperature at high redshift ($z>30$), the 21 cm signal would
come as `emission' if the energy dissipation is efficient. In
Fig.~\ref{fig:KI1} the global HI signal with several magnetic field
strengths are shown. For cases with sufficient magnetic fields, say
$B\gtrsim 0.03$ nG, the signal is always emission against CMB while in
the standard $\Lambda$CDM model the signal would be absorption for the
frequency range of $f_\nu \lesssim 80$ MHz (corresponding to the signal
from redshift $z\gtrsim 20$).

We show the angular power spectrum of the 21 cm brightness temperature
including the PMFs in Fig.~\ref{fig:KI1}b
\citep{2014arXiv1403.2608S}. Here we do not account for any (standard)
heating effects (i.e., UV, X, and L$\alpha$ background emissions) to isolate and
clarify the effects from the PMFs.  On large scales which may be
relevant to SKA 
observations, there are two distinct contributions. One is from the
standard (adiabatic) density fluctuations enhanced by the heating from
the PMFs, and the other is from the PMF induced density fluctuations
dominant on smaller scales
\citep{2006MNRAS.372.1060T,2009ApJ...692..236S}. We can see from the
figure that $B=1$ nG
magnetic fields are marginally within reach for a statistical detection
of the power spectrum. Stacking observing channels in principle will add
more statistical power.

The angular correlation function in real space including the effects
from the PMFs is also studied in \cite{2009JCAP...11..021S}.  The
function exhibits a distinct feature because the PMFs induce early
structure formation and the small scale halos form more compared to the
case in the standard $\Lambda$CDM model The signal from primordial
magnetic fields shows oscillatory feature contrary to that in the
standard $\Lambda$CDM since the matter power spectrum induced by the
PMFs is blue and most of halos are formed at the scale close to the
magnetic Jeans' length. It has been argued that $5$ sigma detection of
the $0.5$ nG magnetic fields will be possible with less than one week
integration of SKA observation \citep{2009JCAP...11..021S}.

\section{Bulk flows}

\cite{2010PhRvD..82h3520T} demonstrated the existence of coherent supersonic velocity flows between baryons and dark matter after decoupling at $z\approx1100$. This has the consequence of inhibiting the formation of star forming galaxies in low-mass halos ($M\lesssim10^6{\rm\,M_\odot}$) \citep[e.g.][]{2011MNRAS.412L..40M,2011ApJ...730L...1S}. If there is significant star formation in such halos, dependent upon H$_2$ cooling and the absence of Lyman-Werner background, then the radiation from such galaxies can lead to a significantly enhanced 21cm signal \citep{2012Natur.487...70V,2014MNRAS.437L..36F,2012ApJ...760....3M}. The effect of bulk flows could then have a major effect on the 21cm signal during the period of Lyman alpha coupling and IGM heating.

Although the details of this are still quite uncertain, if this enhancement exists, it opens the possibility of measuring cosmology at $z=20-27$ with SKA. The relative velocity flows trace the sound horizon and so especially enhance the baryon acoustic oscillation feature in the 21cm power spectrum \citep[see e.g.][]{2012ApJ...760....3M}. The BAO signature provides a standard ruler, calibratable with CMB observations, to form an inverse distance ladder stretching from $z=1100$ through $z\sim20$. Such measurements would strongly constrain the parameter space for departures from $\Lambda$CDM, such as early dark energy models \citep[e.g.][]{2006A&A...454...27B,2006JCAP...06..026D}. Imaging the 21cm structures induced by these bulk flows will be possible with both SKA-LOW Phase 1 and SKA2. This will be a truly novel probe of cosmology at high redshift vastly the reducing the possibility that non-standard dynamics could hide in the observational void between low redshift galaxy surveys and the CMB.

\section{Cosmology on ultra-large scales with SKA-Low}

The measurement of very large scales provides an unique way to probe modifications to the standard cosmological model. In particular it is on scales past the matter-radiation equality peak that General Relativistic corrections become important, at scales above $\sim 5$ comoving Gpc/h \citep{2012PhRvD..85b3504J}. Probing this region would allow to check for any inconsistencies in General Relativity. However, current surveys are still far from probing this region - as an example, the BOSS survey only probes scales up to 200 Mpc/h \citep{2012MNRAS.427.3435A}. In fact, the total volume contained at $z<1$ is sufficient to probe only $\sim8$ modes with $k=2\pi/(5{\rm Gpc}/h)$ while at $6<z<30$ almost 200 modes could be constrained reducing cosmic variance.

Moreover, primordial non-Gaussianity will affect the clustering of biased tracers of dark matter, by adding an extra correction $\Delta b_X(z,k)$ to the Gaussian large-scale bias $b_X^G$ of a given biased tracer $X$: $\Delta b_X(z,k)=3[b_X^G(z)-1]\om\ho^2\delta_c/[c^2k^2T(k)D_+(z)]\fnl$. Here, $\om=\ob+\odm$ is the total (baryons plus dark matter) matter fraction, $\ho$ is the Hubble constant, $\delta_c\simeq1.686$ is the critical collapse density contrast of matter, $T(k)$ is the matter transfer function versus the physical wave number $k$, and $D_+(z)$ is the linear growth factor of density perturbations. Attempts at detecting this effect with redshift surveys have led to some constraints on $\fnl$ \citep{2014PhRvD..89b3511G}.

\begin{figure}[!t]
\includegraphics[scale=0.39]{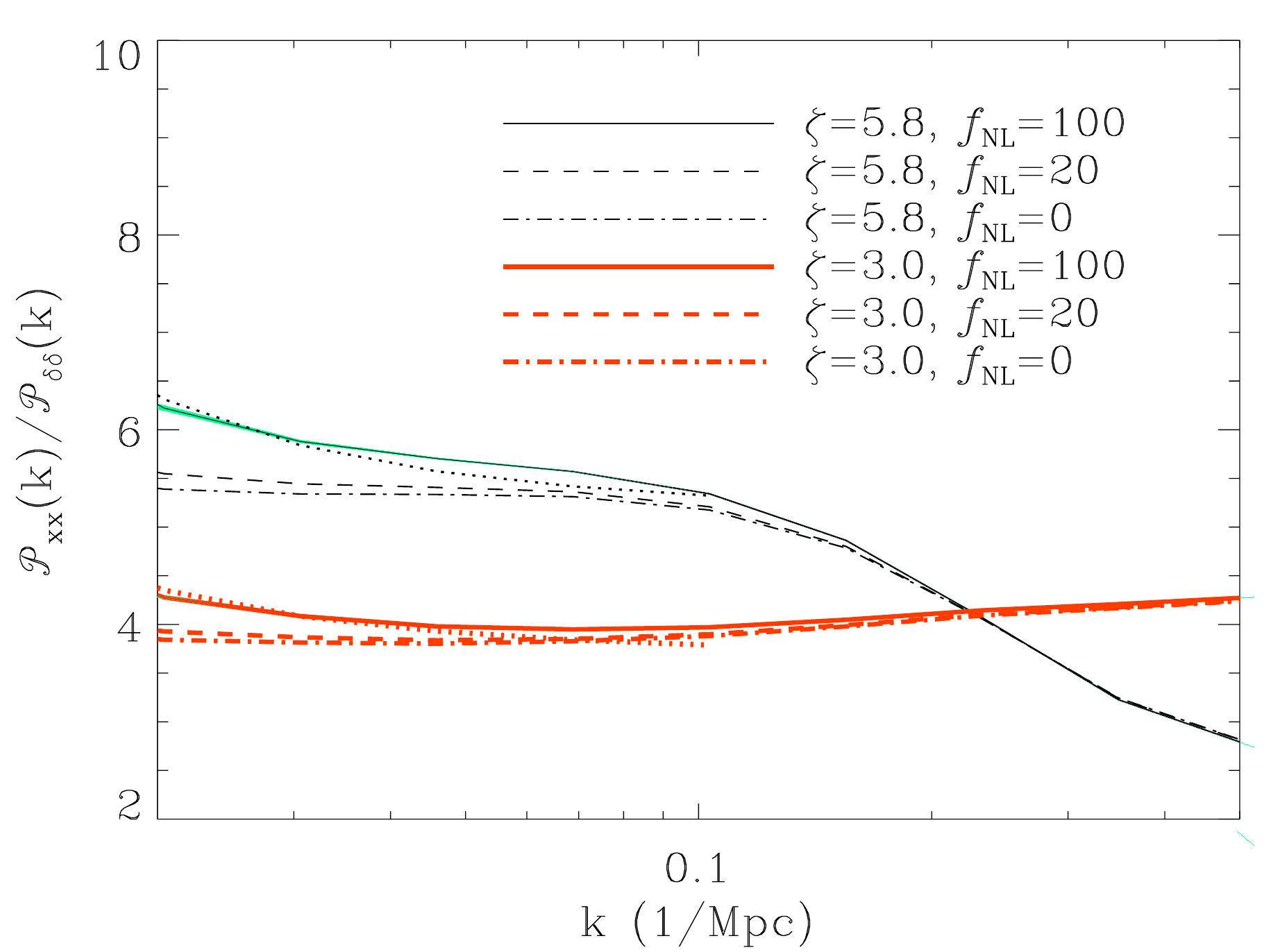}
\includegraphics[scale=0.39]{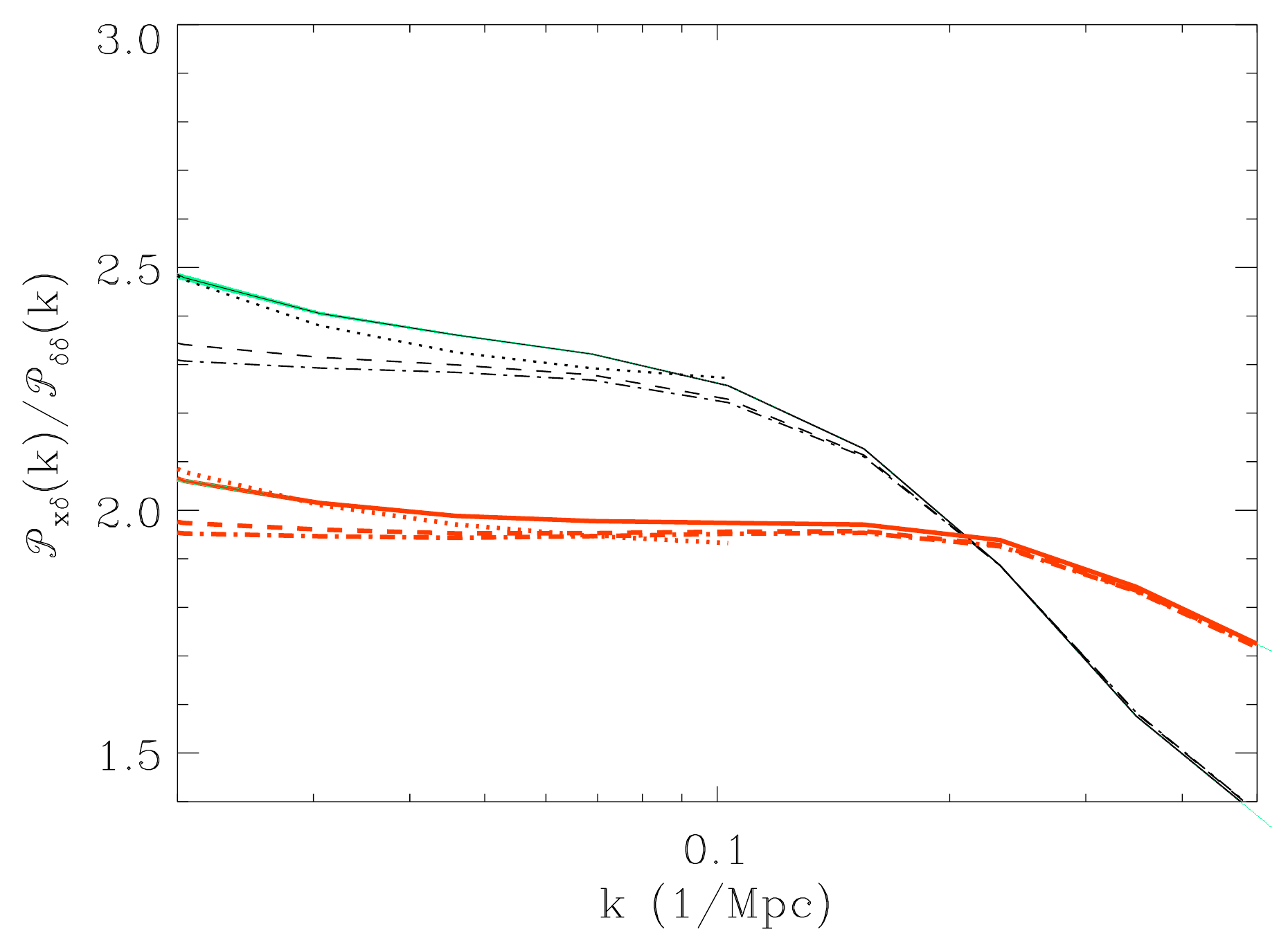}
\caption{Ionization power spectra with non-Gaussianity of the local form from numerical simulations. 
We show $\fnl = (0,20,100)$ (dot-dashed, dashed, solid) for efficiency $\zeta = (5.8, 3.0)$ (thin black, thick red) at $z = 7.5$, where $x_{\rm HI} = (0.50, 0.75)$. 
Analytical fits are in dotted lines.}
\label{figurett}
\end{figure}

By observing at very high redshifts, a low frequency interferometer would be able to probe these large 3-dimensional scales during the Epoch of Reionization and beyond. Although astrophysics will generate model dependent features on the 21cm power spectrum, it is expected that on large enough scales the power spectrum should follow the dark matter one with a different amplitude. For instance, during the epoch of reionization, the ionisation power spectrum should be a biased linear tracer of the dark matter one on scales much larger than the bubble size (see Figure \ref{figurett}. Although the bias itself will depend on the assumed astrophysical model, measurements on these scales will allow to pick any scale dependence generated by primordial non-Gaussianity  or General Relativistic corrections.

\begin{figure}[!t]
\includegraphics[scale=0.38]{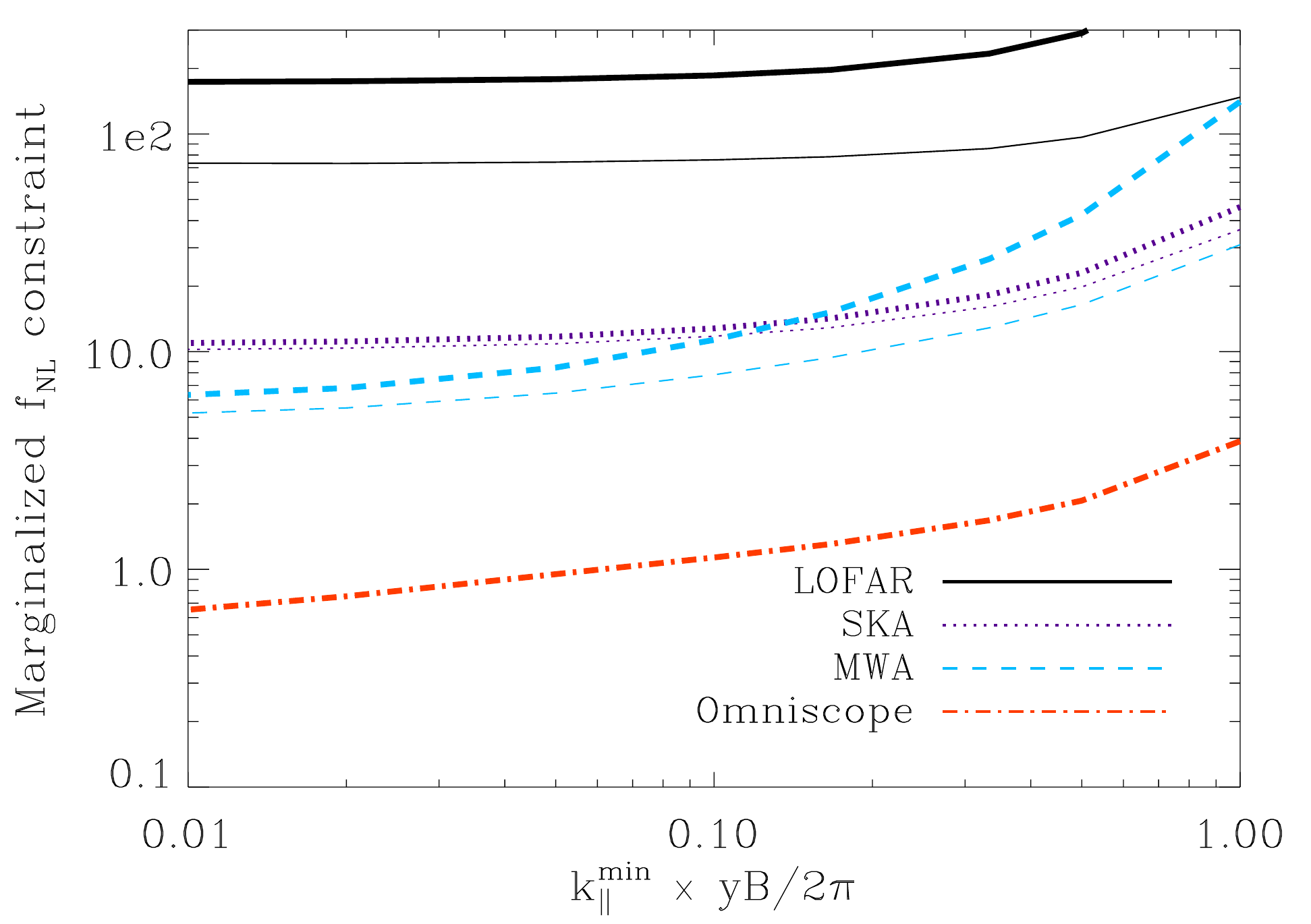}
\includegraphics[scale=0.38]{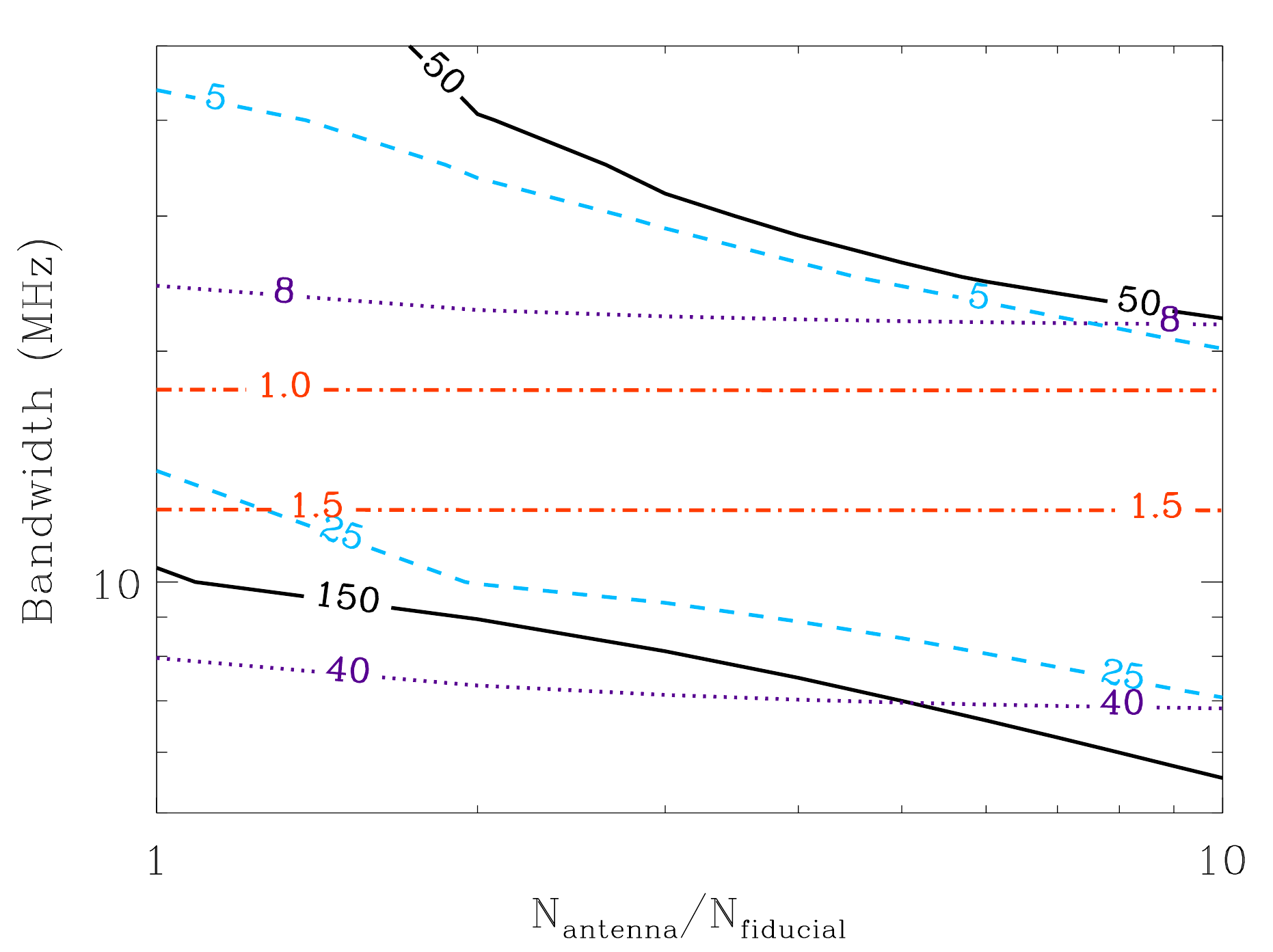}
\caption{{\it Top}: Marginalized $\fnl$ constraints for cases with noise (thick) and without noise (thin), which overlap for Omniscope. We consider a bandwidth of 6 MHz, but assume foregrounds can be removed on scales larger than $k_\parallel = 2\pi/(yB)$. 
{\it Bottom}: Marginalized $\fnl$ constraints as function of bandwidth and number of antennae.
The bandwidth limits the number of modes and largest scale probed along the LOS (via the survey volume $V \propto B$ and $k_\parallel^{\rm min} \propto 1/B$), whereas a larger number of antennae for fixed array density increases the survey resolution and number of perpendicular modes (via $n(u_\perp)$, on large scales $\propto N_{\rm ant}$, and $u_\perp^{\rm max} \propto \sqrt{N_{\rm ant}}$).
The color coding is the same as for the top panel.}
\label{fnl}
\end{figure}

By accessing these large volumes, probes of the high-z 21cm signal will not only measure these large scales but also have enough modes to reduce cosmic variance on the scales of interest.
Having large fields of view will therefore be a key factor for these experiments. The field of view of SKA1-Low ranges from 7 deg$^2$ at 220 MHz ($z\sim 5.5$) to 133 deg$^2$ at 50 MHz ($z\sim 27$), going basically as $(1+z)^2$. These correspond to scales of $\sim 380$ comoving Mpc at $z\sim 5.5$ to $\sim 2.3$ comoving Gpc at $z\sim 27$. As a further example, a fixed redshift bin of 0.1 would evolve on the same range from 50 Mpc to 5.5 Mpc (decreasing with $z$). We see therefore that, although the volumes accessible by SKA1-Low are quite large, especially at very high-z, we would require a telescope with about 100 deg$^2$ at $z\sim 8$ to probe Gpc scales. Something that will probably have to wait for SKA2. Figure \ref{fnl} shows the constraints on the primordial non-Gaussianity parameter for different telescopes.

\section{Cosmic shear and the EoR}
It is possible that the EoR signal could be used to measure weak gravitational lensing.
In \cite{Zahn:2005ap} and \cite{Metcalf:2009}  it was shown that if the EoR is at redshift 
$z \sim 8$ or later, a large radio telescope such as the SKA could measure the lensing convergence power spectrum.  However a very large $f_{\rm sky}$ and a very compact low frequency array was assumed by those authors.  Here the calculation is repeated with parameters that are more consistent with current SKA baseline design.  

The current plans for a 25 square degree survey with SKA1-Low will preclude making competitive measurements of the cosmological parameters through their effects on the weak lensing power-spectrum because of sample variance (this is not true of the SKA1-Mid at lower redshift where the survey area will be much larger).  It still might be possible to map the lensing convergence within the 25 square degree EoR survey area and a mix of wider-shallower and narrower-deeper observing modes can further optimise for weak lensing.  This would allow us to actually ``see'' the distribution of dark matter in a typical region of the sky, something that is only possible with galaxy lensing around very atypical, large galaxy clusters.  This would provide a great opportunity to correlate visible objects with mass and test the dark matter paradigm.

The previously mentioned authors extended the
Fourier-space quadratic estimator technique, which was first developed in 
\cite{Hu:2001tn} for CMB lensing  observations to three dimensional
observables, i.e. the $21$ cm intensity field $I(\theta,z)$.  
The  convergence
estimator and the corresponding variance on the lensing reconstruction are
calculated assuming that the temperature (brightness) distribution is
Gaussian. This will not be strictly true during the EoR, but serves as a reasonable approximation for these purposes. Note that the lensing reconstruction noise contains the thermal noise of the telescope which -assuming a uniform telescope distribution- is calculated using the formula

\begin{equation}
C^{\rm N}_\ell = \frac{(2\pi)^3 T^2_{\rm sys}}{B t_{\rm obs} f^2_{\rm cover} \ell_{\rm max}(\nu)^2} \, ,
\end{equation} where the system temperature $T_{\rm sys}$ at high redshifts
 is dominated by galactic synchrotron radiation and can be approximated by $T_{\rm sys} = 60 \times (\nu/300 \, {\rm MHz})^{-2.55} \, {\rm K}$ \citep{Dewdney:2013}, $B$ is the chosen frequency window, $t_{\rm obs}$ the total observation time, $D_{\rm tel}$ the diameter (maximum baseline) of the core array, $\ell_{\rm max}(\lambda)=2\pi D_{\rm tel}/\lambda$ is the highest multipole that can be measured by the array at frequency $\nu$ (wavelength $\lambda$), and $f_{\rm cover}$ is the total collecting area of the core array $A_{\rm coll}$ divided by $\pi(D_{\rm tel}/2)^2$. 
 
The advantage of 21cm lensing is that one is able to combine
information from multiple redshift slices. In Fourier space, the
temperature fluctuations are divided into perpendicular to the line of
sight wave vectors $\mathbf{k_\perp}=\mathbf{l}/r$, with $r$ the angular diameter distance to the source redshift, and a
discretized version of the parallel wave vector $k_\parallel =
\frac{2\pi}{{\cal L}}j$, where ${\cal L}$ is the depth of the observed
volume. Considering modes with different $j$ independent, an optimal
estimator can be found by combining the individual estimators for
different $j$ modes without mixing them. The three-dimensional lensing reconstruction noise is then found to be  \citep{Zahn:2005ap}\begin{equation}
N(L,\nu) =  \left[\sum_{j=1}^{j_{\rm max}} \frac{1}{L^2}\int \frac{d^2\ell}{(2\pi)^2}  \frac{[\mathbf{l} \cdot \mathbf{L} C_{\ell,j}+\mathbf{L} \cdot (\mathbf{L}-\mathbf{l})
C_{|\ell-L|,j}]^2}{2 C^{\rm tot}_{\ell,j}C^{\rm tot}_{|\mathbf{l}-\mathbf{L}|,j}}\right]^{-1}.
\end{equation}
Here, $C^{\rm tot}_{\ell,j}=C_{\ell,j}+C^{\rm N}_\ell$, where $C_{\ell,j}=[\bar{T}(z)]^2P_{\ell,j}$ with $\bar{T}(z)$ the mean observed brightness temperature at redshift $z$ due to the average HI density and $P_{\ell,j}$ the underlying dark matter power spectrum  \citep{Zahn:2005ap}. 
For SKA1-Low we can consider a $1,000~{\rm hr}$ observation time and we choose $B = 8 \, {\rm MHz}$ and $j_{\rm max} \sim 40$, but with multiple bands $\nu$ that can be stacked to reduce the noise so that $N_L = 1/\displaystyle\sum_{\nu} [N(L,\nu)]^{-1}$. 
 
\begin{figure}[h]
\centerline{
\includegraphics[scale=0.5]{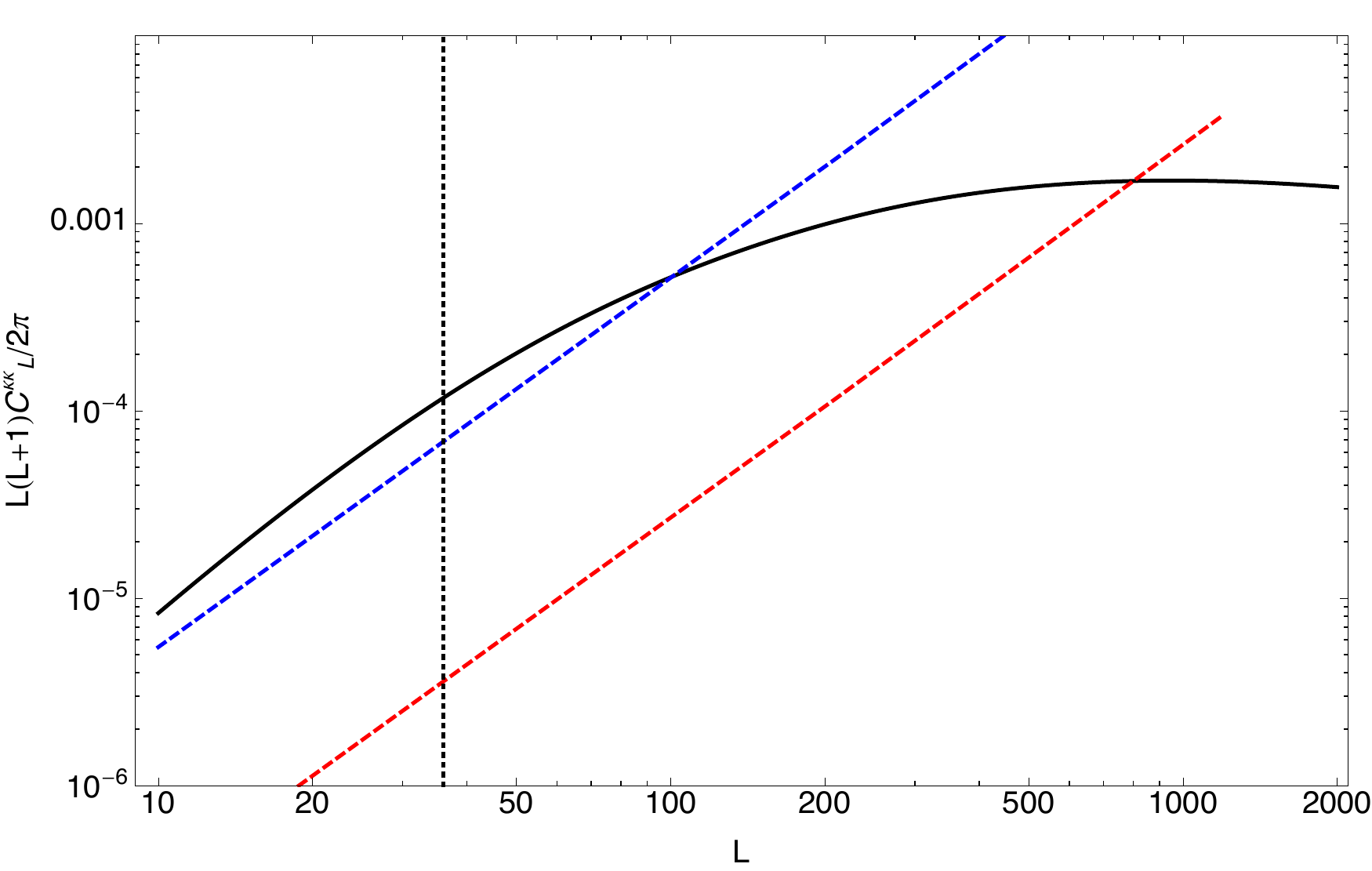}
}
\caption{The lensing convergence field power spectrum, $C^{\kappa \kappa}_L$, for sources at $z=8$ is shown as a solid black line and lensing reconstruction noise $N_L$ as dashed lines.  The blue dashed line is for SKA1-Low with 10 8~MHz frequency bins around $z=8$ spanning the redshift range $z \simeq 6.5-11$.  The red dashed line is for SKA2 and the same frequency bins. The vertical line is approximately the lowest $L$ accessible with a 5-by-5 degree field.  Where the noise curves are below $C^{\kappa \kappa}_L$, typical fluctuations in the lensing deflection should be recoverable in a map. }
\label{fig:CLNL}
\end{figure}

At redshift $z_s \sim 8$, we can assume the SKA1-Low Baseline Design \citep{Dewdney:2013} parameters of $A_{\rm coll} \simeq 0.3  \, {\rm km}^2$ with maximum baseline $D_{\rm tel}=4 \, {\rm km}$, while for SKA we can consider $A_{\rm coll} \simeq 1.2  \, {\rm km}^2$.
The estimated lensing noise is shown in Figure~\ref{fig:CLNL} along with the estimated signal.  
Here $C^{\kappa \kappa}_L$ is the convergence field power spectrum at $z_s=8$ and $N_L$ the lensing reconstruction noise assuming a reionization fraction $f_{\rm HI}=1$.

These results show that it might be possible to map the lensing signal over a range of angular scales.  This measurement greatly benefits from the larger collecting area that will come with SKA2 (we also note that  considering a more compact array, i.e. smaller $D_{\rm tel}$, also improves the signal-to-noise).  Note that the sensitivity of SKA1-LOW is such that larger area surveys at lower integration time do not significantly improve this picture. Although multi-beaming to get larger area at fixed integration time would help with cosmic variance on large scales. For SKA-LOW an optimal observing strategy includes a mix of shallow-wide surveys to beat down cosmic variance on large scales and deep-narrow surveys to measure small scales. The weak lensing power spectrum can be better measured for redshifts after reionization using SKA-Mid and the same 21 cm intensity mapping technique discussed, but over a much larger area of sky \citep{PourtsidouMetcalf:2014}.

\section{Conclusions}

SKA-LOW will provide the first window onto cosmological information from the epoch of reionization and cosmic dawn at $z=6-27$. This makes it unique among the diverse array of future cosmological experiments, which are typically limited to redshifts $z\lesssim3$. This new view of the Universe will test the standard cosmological model deep in the matter dominated regime, where, in principle, we believe we know the evolution of the Universe. The absence of deviations from $\Lambda$CDM would observationally confirm our current assumption that a single cosmological model applies from the CMB to the present. Deviations from $\Lambda$CDM predictions would signal new physics and provide a new way of learning about the Universe.

Precision cosmology with SKA-LOW has potential to be very interesting, but is made difficult by astrophysical contamination. The large volume and number of linear modes of the density field accessible at high redshift will one day lead to a revolution in precision cosmological constraints. SKA-LOW Phase 1 will take the first step in this direction and a mix of shallow-wide and deep-narrow observing fields can help optimise for cosmology; for example matching a narrow-deep 1000hr field over a single field with broad-shallow 10hr fields. The greater sensitivity of SKA2 will push significantly further. The key limitation will be our ability to separate astrophysics the density field to measure cosmological parameters. It is clear that the effort should be made, but our existing understanding of the astrophysics is still too crude to make robust predictions. SKA will contribute to more precise measurements of $\Lambda$CDM parameters, but crucially will test the parameters in a new regime.

Much clearer is the ability of both SKA-Low Phase 1 and 2 to measure the thermal history of the Universe at redshifts $z=6-27$. Never before measured, the thermal history contains information about exotic physics - dark matter physics, primordial magnetic fields, and more. The time dependence and spatial variation of such heating is qualitatively different from that of galaxies providing a clear observational signature accessible to SKA. At the same time, SKA could measure the effect of supersonic relative velocity flows between baryons and dark matter. This probes much the same physics as the CMB and might allow BAO measurements of the sound horizon as a standard ruler to measure the geometry of the Universe.

As SKA1 becomes SKA2, greater sensitivity will allow wider sky surveys to fixed depth for a given integration time. These large volume observations will access super-horizon physics and, with the SKA2, it will be possible to search for relativistic corrections and primordial non-Gaussianity on $\gtrsim$ Gpc scales. Wider sky area and greater angular resolution also improves the ability of SKA to measure weak lensing of the 21cm background. Weak lensing will map the dark matter field in representative patches of the Universe on scales from degrees to tens of arc minutes providing a unique view of the growth of dark matter into the cosmic web.

In this chapter, we have focussed on the main paths to cosmology with SKA-LOW. Many more speculative ideas for cosmology exist. For example, constraining time evolution of the fine structure constant \citep{2007PhRvL..98k1301K} or observing cosmic strong wakes \citep{2010JCAP...12..028B}. 21cm observations over wide sky areas offer a way of probing distinct Hubble volumes in the same way as the CMB. Given the relative infancy of this field, it is likely that new ideas for probing cosmology will be developed during the development of SKA.

It is hopefully clear that SKA-LOW will have something to say on a wide range of cosmological topics. The 21cm signal offers more varied routes to cosmology than traditional probes of large scale structure. This means that the subtleties of extracting precision cosmology from 21cm observations are not fully understood and much work will be needed before SKA. SKA-LOW will transform our view of the astrophysics of the epoch of reionization and cosmic dawn, the same observations will add to our understanding of cosmology and provide new insights into the make up of the early Universe.


\bibliographystyle{apj}
\bibliography{cosmchapterbib}


\end{document}